\documentstyle[epsf]{mn}
 \input epsf.sty
 \newif\ifAMStwofonts
 \ifoldfss

  \ifCUPmtlplainloaded \else
    \NewTextAlphabet{textbfit} {cmbxti10} {}
    \NewTextAlphabet{textbfss} {cmssbx10} {}
    \NewMathAlphabet{mathbfit} {cmbxti10} {} 
    \NewMathAlphabet{mathbfss} {cmssbx10} {} 
  \fi
  \ifAMStwofonts
    \ifCUPmtlplainloaded \else
      \NewSymbolFont{upmath} {eurm10}
      \NewSymbolFont{AMSa} {msam10}
      \NewMathSymbol{\upi}     {0}{upmath}{19}
      \NewMathSymbol{\umu}     {0}{upmath}{16}
      \NewMathSymbol{\upartial}{0}{upmath}{40}
      \NewMathSymbol{\leqslant}{3}{AMSa}{36}
      \NewMathSymbol{\geqslant}{3}{AMSa}{3E}

    \fi
  \fi
 \fi 

 \ifnfssone
  \newmathalphabet{\mathit}
  \addtoversion{normal}{\mathit}{cmr}{m}{it}
  \addtoversion{bold}{\mathit}{cmr}{bx}{it}
  \newmathalphabet{\mathbfit} 
  \addtoversion{normal}{\mathbfit}{cmr}{bx}{it}
  \addtoversion{bold}{\mathbfit}{cmr}{bx}{it}
  \newmathalphabet{\mathbfss} 
  \addtoversion{normal}{\mathbfss}{cmss}{bx}{n}
  \addtoversion{bold}{\mathbfss}{cmss}{bx}{n}
  \ifAMStwofonts
    \ifCUPmtlplainloaded \else
      \UseAMStwoboldmath
      \makeatletter
      \new@mathgroup\upmath@group
      \define@mathgroup\mv@normal\upmath@group{eur}{m}{n}
      \define@mathgroup\mv@bold\upmath@group{eur}{b}{n}
      \edef\UPM{\hexnumber\upmath@group}
      \new@mathgroup\amsa@group
      \define@mathgroup\mv@normal\amsa@group{msa}{m}{n}
      \define@mathgroup\mv@bold\amsa@group{msa}{m}{n}
      \edef\AMSa{\hexnumber\amsa@group}
      \makeatother
      \mathchardef\upi="0\UPM19
      \mathchardef\umu="0\UPM16
      \mathchardef\upartial="0\UPM40
      \mathchardef\leqslant="3\AMSa36
      \mathchardef\geqslant="3\AMSa3E
    \fi
  \fi
 \fi 

 \ifnfsstwo
  \DeclareMathAlphabet{\mathbfit}{OT1}{cmr}{bx}{it}
  \SetMathAlphabet\mathbfit{bold}{OT1}{cmr}{bx}{it}
  \DeclareMathAlphabet{\mathbfss}{OT1}{cmss}{bx}{n}
  \SetMathAlphabet\mathbfss{bold}{OT1}{cmss}{bx}{n}
  \ifAMStwofonts
    \ifCUPmtlplainloaded \else
      \DeclareSymbolFont{UPM}{U}{eur}{m}{n}
      \SetSymbolFont{UPM}{bold}{U}{eur}{b}{n}
      \DeclareSymbolFont{AMSa}{U}{msa}{m}{n}
      \DeclareMathSymbol{\upi}{0}{UPM}{"19}
      \DeclareMathSymbol{\umu}{0}{UPM}{"16}
      \DeclareMathSymbol{\upartial}{0}{UPM}{"40}
      \DeclareMathSymbol{\leqslant}{3}{AMSa}{"36}
      \DeclareMathSymbol{\geqslant}{3}{AMSa}{"3E}
    \fi
  \fi
 \fi 

 \ifCUPmtlplainloaded \else
  \ifAMStwofonts \else 
    \def\upi{\pi}
    \def\umu{\mu}
    \def\upartial{\partial}
  \fi
 \fi

 \title[Long Term Light Curve of A0620-00]
  {Regularities in the Long Term Optical Light Curve of the Black
  Hole Candidate Binary A0620-00 (V616 Mon)}

 \author[E.M. Leibowitz, S. Hemar and M. Orio]
   {Elia M. Leibowitz$^1$, Shirley Hemar$^1$ and Marina Orio$^2$\\
  $^1$School of Physics and Astronomy and the Wise Observatory,
 Raymond and Beverly Sackler Faculty of Exact Sciences,\\
 Tel-Aviv University, Tel Aviv, 69978, Israel\\
 $^2$Osservatorio Astronomico di Torino, Italy\\}

 \pagerange{\pageref{firstpage}--\pageref{lastpage}}
 \pubyear{1997}

 \def\LaTeX{L\kern-.36em\raise.3ex\hbox{a}\kern-.15em
    T\kern-.1667em\lower.7ex\hbox{E}\kern-.125emX}

\begin{document}

 \label{firstpage}

 \maketitle

 \begin{abstract}
 We have monitored the R and I magnitude of the black hole candidate system
 A0620-00 (V616 Mon) in the years 1991-1995 at the Wise Observatory.
 Combining our data with some additional measurements, we analyze a sparsely
 covered 7 year light curve of the star. We find that the average R-band
 magnitude is varying on a time scale of a few hundreds days, with a
 peak to peak amplitude of 0.3 mag.

 The two maxima in the well known double hump binary cycle, as well as one
 of the minima between them, vary by a few percent relative to one another,
 in a seemingly random way. One maximum is on the average higher by 0.05 mag.
 than the other. The depth of the second minimum is varying with
 significantly higher amplitude, in clear correlation with the long term
 variability of the mean magnitude of the system. It is shallower than the
 other minimum at times of maximum light. It deepens when the system
 brightness declines, and it becomes the deeper among the two minima at
 times of minimum system light.

 According to the commonly acceptable phasing of the binary cycle, the
 systematically varying minimum corresponds to inferior conjunction of the
 red dwarf. We cannot suggest any simple geometrical model for explaining
 the regularities that we find in the long term photometric behaviour of the
 V616 Mon binary system.
 \end{abstract}


 \section{Introduction}
 The X-ray nova A0620-00 was discovered on August 3, 1975 (Elvis et al. 1975)
 during a routine monitoring of the Milky Way with the Ariel V Sky Survey
 Experiment (SSE). The optical counterpart of the X-ray source was identified
 as the star V616 Mon at the 1950 coordinates: RA=$6^{h}20^{m}11^{s}$.2
 Dec=-$0^{o}$19'10'' (Boley et al. 1976). A previous outburst of the source
 was discovered on Harvard photographic plates taken in 1917 (Eachus et al.
 1976).

 Spectroscopic as well as photometric measurements show that the star is a
 compact binary with a period $P_{B}$=7.75234 hr (McClintock and Remillard
 1986). The secondary component is a K5-K7 red dwarf (Oke 1977). From the
 radial velocity curve of the secondary, a mass function $f(M)=3.18\pm0.16
 M_{\odot}$ can be derived (McClintock and Remillard 1986). If the mass ratio
 does not exceed $q=M_{1}/M_{2}=10.6$ and an upper limit of $0.8 M_{\odot}$
 is placed on the mass of the companion star, a lower limit of $M_{1}=4.16
 M_{\odot}$ can be determined for the mass of the compact object (Haswell et
 al. 1993). As this value is above the theoretical upper limit for the mass
 of a neutron star, this star is considered a black hole candidate (BHC).

 A0620-00 is the first X-ray nova that was identified as a BHC, and it is
 considered a prototype of the small but growing class of similar Galactic
 objects. As the oldest member in the class of BHC/X-ray novae and the
 optically brightest in its quiescence state, it is also probably the most
 thoroughly studied system of its kind, particularly in optical photometry.
 The photometric binary cycle of the system has a clear double hump structure
 (McClintock and Remillard 1986), which is commonly believed to be the result
 of the ellipsoidal effect. The Roche lobe filling secondary is distorted
 into an ellipsoidal shape by the strong gravitational pull of its companion,
 the black hole. As its spin rate is tidally locked with the orbital
 revolution, it reveals its extended dimension to an observer on Earth twice
 during each orbital cycle, at the two phases of maximum light in the
 photometric cycle. Minima in the photometric binary light curve (LC) are
 observed when one of the narrow ends of the ellipsoid is pointing at the
 observer. McClintock and Remillard (1986) determined the epoch of minimum
 light in the binary LC at JD 2,445,477.827. From radial velocity
 measurements they have also established that this is the phase of inferior
 conjunction of the compact object, namely, when the black hole is in front
 of the secondary, with respect to the observer.

 There are scattered reports in the literature on variations in the
 average magnitude of A0620-00 (McClintock and Remillard 1986, Haswell 1992).
 Haswell et al. (1993) and Bartolini et al. (1990) report about variations
 in the structure of the binary LC. In particular the relative height
 between the two maxima in the double humped LC seems to be different in
 different epochs, as well as the relative depth of the two minima.

 The mass function of A0620-00, derived from the radial velocity curve
 establishes a lower limit to the mass of the compact star. However, in order
 to better constrain the possible mass value of the black hole candidate, one
 must have an estimate for the inclination angle {\em i} of the system.
 This may be achieved by analyzing the LC of the system (Haswell 1995,
 Charles 1995). Thus, a detailed study of the optical photometric behavior
 of A0620-00 seems to be an important avenue for better understanding this
 prototype BHC. This was our motivation for initiating at the Wise
 Observatory (WO) a program of long term monitoring of the photometric
 behavior of this star. Here we report the results of this 5 year program.
 We analyze our own data, as well as some additional relevant data, obtained
 by other observers, as described in the next section.

 \section{Observations}
 \subsection{Photometry at the Wise Observatory}
 Photometric observations of A0620-00 (V616 Mon) were performed at WO from
 January 1991 to November 1995. A journal of the observations is presented in
 Table 1. Runs 1 to 17 were performed with the observatory 320x520 RCA CCD
 camera. From Run 18 on we used the newer Tektronix 1024x1024 CCD camera.
 Detailed description of the WO optics and measuring instruments is given by
 Kaspi et al. (1995). Most of our observations (Runs 1-5, 8-21) were made
 through a Cousins R band filter, and some with an I filter (Runs 1-7).
 The typical integration time was 6 to 8 minutes.

 All of our frames were subjected to the usual data reduction procedure of
 bias subtraction and division by flat field frames, taken each night in the
 corresponding filters. Using the program DAOPHOT (Stetson 1987) we then
 performed aperture photometry, deriving from each frame instrumental
 magnitudes of A0620-00, as well as of a few reference stars in its
 neighborhood. We then used the Wise Observatory program DAOSTAT (Netzer et
 al. 1996) to calculate the magnitude of the object, relative to some mean
 magnitude of a group of 10 reference stars that proved to be non-variable
 during the course of our observations. The observational error in the value
 of the relative magnitudes is $\sim.02$ mag, while the error in the
 absolute value is around 0.1 mag.

 \begin{table}
 \caption{Observation Journal from Wise Observatory}
 \begin{center}
 \begin{tabular}{@{}lcccc@{}}

 Observing  & UT Date & Start Time    & Run Time & No. of\\
   Run      &         &JD 2,440,000+  &  (hours) & Points \\
 \\
 1 & 08 Jan. 91 & 8265.23 & 6.02 & 18 \\
 2 & 10 Jan. 91 & 8267.19 & 7.63 & 39 \\
 3 & 05 Feb. 91 & 8293.23 & 5.14 & 21 \\
 4 & 08 Feb. 91 & 8296.21 & 5.57 & 21 \\
 5 & 17 Feb. 91 & 8305.22 & 4.61 & 19 \\
 6 & 27 Jan. 93 & 9015.21 & 5.76 & 24 \\
 7 & 30 Jan. 93 & 9018.19 & 6.00 & 25 \\
 8 & 06 Nov. 93 & 9298.41 & 5.30 & 62 \\
 9 & 10 Nov. 93 & 9302.51 & 2.94 & 27 \\
 10 & 22 Nov. 93 & 9314.34 & 4.78 & 35 \\
 11 & 23 Nov. 93 & 9315.35 & 4.48 & 29 \\
 12 & 08 Dec. 93 & 9330.30 & 5.22 & 38 \\
 13 & 15 Dec. 93 & 9337.32 & 4.67 & 36 \\
 14 & 08 Jan. 94 & 9361.21 & 8.18 & 47 \\
 15 & 09 Jan. 94 & 9362.19 & 8.91 & 84 \\
 16 & 19 Jan. 94 & 9372.27 & 6.36 & 42 \\
 17 & 20 Jan. 94 & 9373.23 & 7.33 & 56 \\
 18 & 09 Feb. 94 & 9393.30 & 3.75 & 32 \\
 19 & 10 Feb. 94 & 9394.21 & 5.03 & 31 \\
 20 & 09 Mar. 94 & 9421.20 & 3.91 & 26 \\
 21 & 06 Dec. 94 & 9693.47 & 4.02 & 35 \\
 22 & 07 Dec. 94 & 9694.43 & 4.92 & 47 \\
 23 & 08 Dec. 94 & 9695.43 & 4.88 & 34 \\
 24 & 09 Dec. 94 & 9696.44 & 4.65 & 34 \\
 25 & 10 Dec. 94 & 9697.41 & 4.77 & 29 \\
 26 & 05 Jan. 95 & 9723.24 & 7.73 & 56 \\
 27 & 09 Feb. 95 & 9758.22 & 5.85 & 55 \\
 28 & 19 Nov. 95 & 10041.41 & 5.16 & 31 \\
 \end{tabular}
 \end{center}
 \end{table}

 \subsection{Additional Observations}
 In this work we analyze in addition to our own data also a few photometric
 results obtained by L. Solmi with the CCD camera attached to the 1.52 m
 telescope of the Bologna University at Loiano, Bologna (Solmi 1989).
 We were able to combine L. Solmi's R data with our own, by finding one
 standard star that was used as a reference star in both sets of observations.
 Based on the measurements of this star, we introduced a shift of -0.0962 mag.
 into Solmi's values and thus obtained a consistent 7 year R LC of A0620-00.
 This LC is analyzed in Sections 3 and 4.

 \section{Analysis}
 \subsection{Long Term Variations}
 Figure 1 shows our 7-year R-band light curve of V616 Mon, with individual
 measurements shown as dots. Squares denote mean magnitudes of 11 subgroups
 of observations. Each subgroup consists of close consecutive measurements
 that together cover well a complete cycle of the binary periodicity.
 The time interval spanned by the observations in each group is detailed
 in Table 2, the largest being 41 days. The mean value displayed is the
 free term in a 3 harmonics Fourier expansion of the corresponding observed
 magnitudes around the binary period $P_{B}$.

 \begin{figure}
 \centerline{\epsfxsize=3.0in}
 \caption{R band light curve of A0620-00 from Feb 89 to Nov 95. The dots
 denote individual measurements. The squares denote the mean magnitude value
 of 11 subgroups of observations.}
 \end{figure}

 \begin{table}
 \caption{11 data groups of R band observations used in this discussion}
 \begin{center}
 \begin{tabular}{@{}lccc@{}}
 Data  & JD 2,440,000+ & Time Span & No. of \\
 Group &               &   (days)  & Points \\
 \\
 1 &     8230    &  1 & 20\\
 2 & 7567 - 7577 & 11 & 38\\
 3 & 8265 - 8305 & 41 & 29\\
 4 & 9298 - 9302 &  5 & 89\\
 5 & 9314 - 9337 & 24 & 138\\
 6 & 9361 - 9362 &  2 & 131\\
 7 & 9372 - 9373 &  2 & 98\\
 8 & 9393 - 9421 & 29 & 89\\
 9 & 9693 - 9697 &  5 & 179\\
 10 & 9723 - 9758 & 36 & 111\\
 11 & 10041 & 1 & 31\\

 \end{tabular}
 \end{center}
 \end{table}

 A power spectrum analysis (Scargle 1982) was performed over the entire R
 band  data set. In its high frequency end, the power spectrum is dominated
 by a very high peak corresponding to the periodicity 0.1615080 days. Twice
 this value should be the photometric binary period of the system $P_{B}=
 0.3230160 day = 7.75238 hour$. Epoch of phase 0, i.e. of minimum M2 (see
 Section 3.2), is at JD 2450000.025. In the long period end of the power
 spectrum there is a high peak at the frequency corresponding to a period
 of 255 days. The power spectrum of just the 11 mean magnitude values has
 also its highest peak at the same frequency.

 The peaks around the 255 d periodicity in both power spectra, that of the
 entire data set and that of the 11 point LC, are statistically {\em not}
 significant. There is, however, hardly any doubt that the mean brightness
 of the star varies with an amplitude of $\sim0.3$ magnitudes on a time
 scale of hundreds of days.  The change in the general brightness of the
 star is also mentioned by McClintock and Remillard (1986) and by
 Haswell (1992).

 \subsection{The Structure of the Binary Light Curve}
 For each one of the 11 groups of data points discussed in the previous
 section we calculated a mean binary LC by least squares fitting to the
 data the first 3 harmonics of the Fourier expansion around the known binary
 period. All 11 light curves show the double hump structure with variations
 in the amplitudes of the extrema from one LC to another. Three examples are
 shown in Figure 2a-c. The typical, smooth LC of A0620-00, as shown in
 Figure 2d, is characterized by two minima - M2 and M1 around phases 0.0
 and 0.5, and two maxima X1 and X2 near phases 0.25 and 0.75. Phase 0 here
 and in all other binary LCs shown in this work is at JD2445477.827, as in
 McClintock and Remillard 1986.

 For each of the 11 LCs established from our observed data, we determined the
 magnitude of the star at X1, M1, X2 and M2, by considering the 4 extremum
 points in the fitted smooth curve. The mean magnitude value M was determined
 for each LC by the free term in the Fourier expansion of the fitted curve.

 \begin{figure}
 \centerline{\epsfxsize=3.0in}
 \caption{(a, b, c) Light curves of A0620-00 at different times. The smooth
 line represents a 3 harmonics least squares fit of the known orbital period.
 (d) Scheme of a typical light curve of the star. X1, M1, X2, M2 denote the
 four extremum points of the LC. M denotes the mean magnitude of the star at
 that time.}
 \end{figure}

 As evident from Figure 2, the structure of the binary LC and the mean
 magnitude of the star vary from one data group to another. To isolate the
 structural effect we consider magnitude differences between extremum points
 in a given binary LC. Figure 3a is a plot of the magnitude difference
 X1-X2 (* symbol) and of M1-M2 (diamond symbol) vs. the system mean
 magnitude M. Figure 3b is a plot of the magnitude difference X1-M1
 (triangle symbol) and of X2-M2 (X symbol) vs. the system mean magnitude M.
 The solid and dotted straight lines are the linear regression lines of the
 corresponding sets of points.

 The error bars in the figures were determined by the bootstrap method
 (Efron and Tibshirani 1993). For each LC we calculated all the differences
 $d_{i}$ between observed magnitudes and the corresponding values on the
 smooth curve. We then added to each magnitude on the fitted curve one
 number, drawn randomly from the ensemble of the $d_{i}$ values (with
 repetitions). We thus created a new pseudo-observed binary LC, to which
 we fitted by least squares a smooth LC of 3 Fourier harmonics. For this LC
 we found, as before, the magnitude value of the 4 extremum points. Repeating
 the procedure 1000 times we obtained 1000 values of magnitudes for each
 extremum point. The error bars are the half width at 0.04 of the maximum
 in the histogram of the corresponding parameter.

 \begin{figure}
 \centerline{\epsfxsize=3.0in}
 \caption{a) Magnitude differences X1-X2 (* symbol) and M1-M2 (diamonds) vs.
 mean magnitude (M) for the 11 groups of R data. The dotted line represents
 a least squares linear fit to the X1-X2 points and the solid line represents
 a least squares linear fit to the M1-M2 points. (b) Magnitude differences
 X1-M1 (triangles) and X2-M2 (x symbol) vs. mean magnitude (M) for the 11
 groups of R data. The solid line represents a least squares linear fit to
 the X1-M1 points and the dotted line represents a least squares linear fit
 to the X2-M2 points.}
 \end{figure}

 \subsection{Photometric Correlation}
 In Figure 3a one can see that although the X1-X2 magnitude difference
 (* symbol) varies by up to 0.1 mag, there is no dependence of this variation
 on M. The mean value of this parameter is around 0.05 mag. On the other
 hand, the M1-M2 parameter (diamond symbol) appears to be correlated with
 the system mean magnitude M. This is also borne out by statistical tests.
 The dotted straight line in Figure 3a is a regression line fitted to the
 X1-X2 points by least squares. The slope of this line is 0.043. By the
 bootstrap method (Efron and Tibshirani 1993), on a pseudo-sample of 1000,
 we find a 60\% confidence interval around this value of $\pm0.06$. Thus
 the slope of this line is consistent with being zero. On the other hand,
 the slope of the solid, regression line of the M1-M2 parameter values is
 0.381. With the bootstrap method we find that the probability that the
 slope is 0 or negative is not larger than 0.2\%, i.e. the slope is
 significantly different from 0.

 The slope of the solid regression line in Figure 3a is admmitedly very much
 dependent on the value of one single point. When we remove from the data
 the diamond point at the extreme right end of the figure, the slope of
 the regression line of the remaining 10 points is no longer significantly
 different from 0. We do believe however, that the correlation presented in
 the solid line in Figure 3a is nevertheless significant. The position of
 that crucial point in the plane of the graph is well established, as evident
 from the very small error bar around it. Secondly, we made a further check
 by deviding all the observed measurements into 12 subgroups rather than to
 the 11 groups as described in Section 3.2. We plotted the
 12 M1-M2 values vs. M as in Figure 3a, and computed the corresponding
 regression line. Its slope is similar to that of the solid line in
 Figure 3a, and it does not depend critically on any single point among
 the 12 in that plot. As a third test for the significance of the slope of
 the solid line in Figure 3a we performed a second, different bootstrap
 analysis, by randomly reshuffling the 11 M1-M2 values among the 11 mean
 magnitude (M) values. We computed the regression lines of the 1000 different
 distributions of 11 (M,M1-M2) points so obtained. In only 18 cases out of
 the 1000, the slope was found equal or larger than the value in the real,
 observed data.

 In Figure 3b we plot two other magnitude differences among the 4 extremum
 points of the binary LC. The triangle symbol denotes the X1-M1 difference
 and the X symbol represents the X2-M2 difference. Here we see again that one
 parameter is independent of M, while the other one is correlated with it.
 The slope of the dotted straight line in Figure 3b, the regression line of
 the X2-M2 parameter values, is 0.066. With a 60\% confidence interval of
 $\pm0.07$ it is consistent with being 0. The mean value of the X2-M2
 magnitude difference is -0.15. On the other hand, the slope of the solid
 regression line of the X1-M1 parameter, is -0.272. With the bootstrap we
 find that the probability that the slope of the true line of regression
 is 0 or positive is smaller than 5\%. Here again we performed in addition
 the other two checks described in the previous paragraph. They too show
 that the slope of the solid regression line in Figure 3b is significantly
 {\em different} from 0.

 \begin{table}
 \caption{The five groups of data points}
 \begin{center}
 \begin{tabular}{@{}cc@{}}
 Mean      & No. of \\
 Magnitude & Data Points\\
 \\
 17.00     & 268\\
 17.09     & 142\\
 17.21     & 118\\
 17.27     & 445\\
 17.33     & 20\\
 \end{tabular}
 \end{center}
 \end{table}

 In order to further check this correlation with a better signal to noise
 ratio, we regrouped all our single-night light curves into 5 new subgroups.
 These are defined according to the mean magnitude value of each LC rather
 than by the time of observations. Table 4 presents the central magnitude
 of each of the 5 bins that we have defined on the magnitude axis, as well
 as the number of observed points in each bin. For each bin we fitted by
 least squares the first 3 harmonics of a Fourier series, expanded around
 the fundamental periodicity $P_{B}$ of the binary cycle. As before, the
 free term in this expansion was taken as the mean magnitude of the system,
 and the 4 extremum points X1, M1, X2 and M2 were determined.

 In Figure 4 we draw all the 5 fitted LCs together, where we assign the
 magnitude value 0 to all X1 maxima. One can clearly see that the resulting
 variation in the magnitude value of minimum M1 is significantly larger than
 the variations in the magnitude value of the other two extremum points, X2
 and M2. The systematic difference between the heights of the 2 maximum
 points is also apparent in this presentation.

 \begin{figure}
 \centerline{\epsfxsize=3.0in}
 \caption{The 5 light curves of data regrouped according to the mean
 magnitude M. The magnitude of maximum X1 in all LCs is set to 0. Minimum M1
 (at phase $\sim0.5$) varies with a considerably larger amplitude than the
 other two extrmum points X2 and M2.}
 \end{figure}

 Figure 5a is a plot of the magnitude differences X1-M1 and X2-M2 of the LCs
 shown in Figure 4, as functions of M, the mean magnitude of the system. The
 straight lines are linear regression lines of the corresponding observed
 points. As in Figure 3a, we see here, that X1-X2 is independent of M, while
 M1-M2 is very much correlated with it. Figure 5b is the corresponding plot
 for the magnitude differences X1-M1 and X2-M2. Again we see, as in Figure
 3b, that X2-M2 is independent of M while X1-M1 is clearly correlated with it.

 \begin{figure}
 \centerline{\epsfxsize=3.0in}
 \caption{Magnitude differences (Upper panel: X1-X2 and M1-M2, Lower panel:
 X1-M1 and X2-M2) vs. mean magnitude for the 5 subsets of R data, grouped by
 magnitude.}
 \end{figure}

 Combining the results presented in Figures 3a and 3b, or in Figures 5a and
 5b, we can arrive at the following conclusion: Among the 4 extremum points
 of the binary light curve of A0620-00, the relative magnitude of three of
 them, namely X1, X2 and M2 (with respect to the mean system magnitude or
 with respect to one another) are independent of M, the mean magnitude of
 the system. The relative depth of minimum M1, on the other hand, is
 correlated with M. It is deepest when the system is faint, and it becomes
 shallower as the system brightens. Figure 5 shows that while the mean
 magnitude of the system varies by 0.3 mag., which is the range of its long
 term brightness variations, the mean depth of minimum M1 varies by
 $\sim0.12$ mag.

 \section{Discussion}
 The BHC system A0620-00 (V616 Mon) is found to vary in its optical
 brightness on a time scale of hundreds of days. This is in addition to its
 well known photometric variation with the binary periodicity of the system.
 The amplitude of the long term variation is $\sim0.3$ mag. It is possibly
 periodic with a period of $P_{L}$=255 d, but this could not be established
 at any statistically significant level from the data at our disposal.

 In 11 different R band binary LCs of the system that were observed over 7
 years, we find systematic variations in the structure that are correlated
 with variations in the mean magnitude of the system. The most apparent
 variations are in the relative magnitude of the 4 extremum points that
 characterize the photometric binary cycle. We find that the relative
 magnitudes of the 3 extremum points X1, X2 and M2 do vary among the
 different LCs. Some of these variations are due to observational
 uncertainties in the determination of the respective magnitude values,
 but some of them seem to be real. The standard deviation of the magnitude
 of these 3 extremum points, relative to the mean system magnitude, are
 0.0255, 0.0266 and 0.0280 mag., respectively. There is no apparent
 correlation of this variation with the mean magnitude of the system.

 The behavior of minimum M1 is quite different. The standard deviation of
 its depth, relative to the mean system magnitude, is 0.0314 mag.,
 significantly larger than the other 3 extremum points. The variations of
 this minimum are well correlated with the system mean magnitude. The
 correlation is in the sense that this minimum is deepest when the system
 is faint, and it becomes shallower as the system brightens. In quantitative
 terms, as the R magnitude of the system varies by 0.30 mag. the relative
 depth of the M1 minimum is varying by $\sim0.12$ mag.

 Variations in the structure of a double hump binary LC, similar to those
 in V616 Mon reported here, have been observed in the massive X-ray
 binary LMC X-4 (Heemskerk and van Paradijs, 1989 - HvP). In that system,
 much like in ours, there are noticeable variations in the relative height
 of the two maxima, as well as in the relative depth of the two minima, with
 one minimum showing much larger variations than the other. In LMC X-4, the
 variations are found to be correlated with the 30 d periodicity of the
 X-ray on/off cycle of that system. HvP interpreted the observed structural
 variations on the basis of a precessing disc model. With a combination of
 a tidal and rotational distorted secondary, X-ray heating of the secondary
 surface, and a luminous precessing disc, they were able to reproduce rather
 faithfully the varying structure of the binary optical LC, at all phases
 of the 30 d periodicity.

 It is tempting to suggest a similar model for the temporal behaviour of the
 optical LC of V616 Mon. There is however one fundamental difference that
 severely harms the analogy between these two cases. In LMC X-4, the minimum
 in the binary LC that varies with the larger amplitude is the one at binary
 phase 0.5, which according to HvP corresponds to inferior conjunction of
 the X-ray source in the system (HvP Figures 9 and 10). This is the binary
 phase at which the illuminated hemisphere of the secondary star is facing
 the observer. This is therefore the phase at which variations in the X-ray
 illumination, due to disc precession, are mostly reflected in the optical
 radiation.

 In V616 Mon the situation is different. Comparing phases of the photometric
 data with spectroscopic radial velocity measurements, McClintock and
 Remillard (1986) determined that the maximum that we denote X2 is the phase
 of maximum radial velocity of the red dwarf in the binary system. The
 varying minimum M1 is accordingly at the phase of inferior conjunction of
 the red dwarf. At this phase the red dwarf is located between the observer
 and the compact object, with its non illuminated hemisphere in the
 direction of the observer. This is the phase when variations in the
 illumination of the secondary are hardly affecting the binary optical LC.
 It therefore seems to us that if no discrepancy is found in the commonly
 adopted relative phasing of the LC and the radial velocity curve, a simple
 geometrical model, based on ellipsoidal and reflection effects, cannot
 account for the photometric long term behavior of A0620-00.

 Needless to say, further photometric monitoring of the system and/or
 publication of additional LCs that are no doubt still in the possession of
 a few observers of this star, are crucial for a proper understanding of
 this prototype BHC.

 \section{Acknoledgements}
 We are grateful to Alon Retter, Haim Mendelson and John Dann for their help
 with the Wise Observatory observations. We also thank Phil Charles for
 some useful comments. Research work at the Wise Observatory is supported
 by the Foundation  for Basic Research of the Israel Academy of Sciences.

 
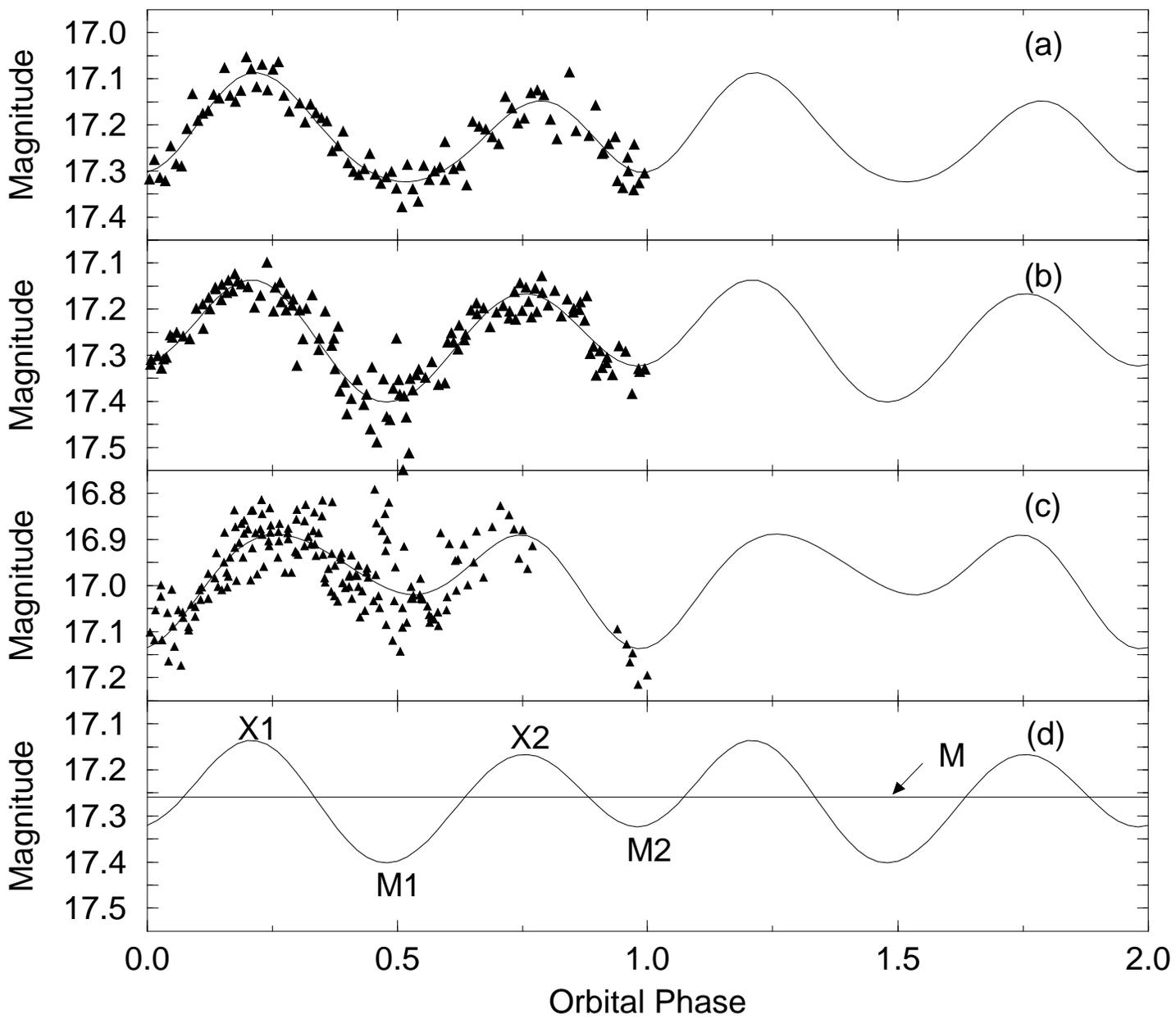
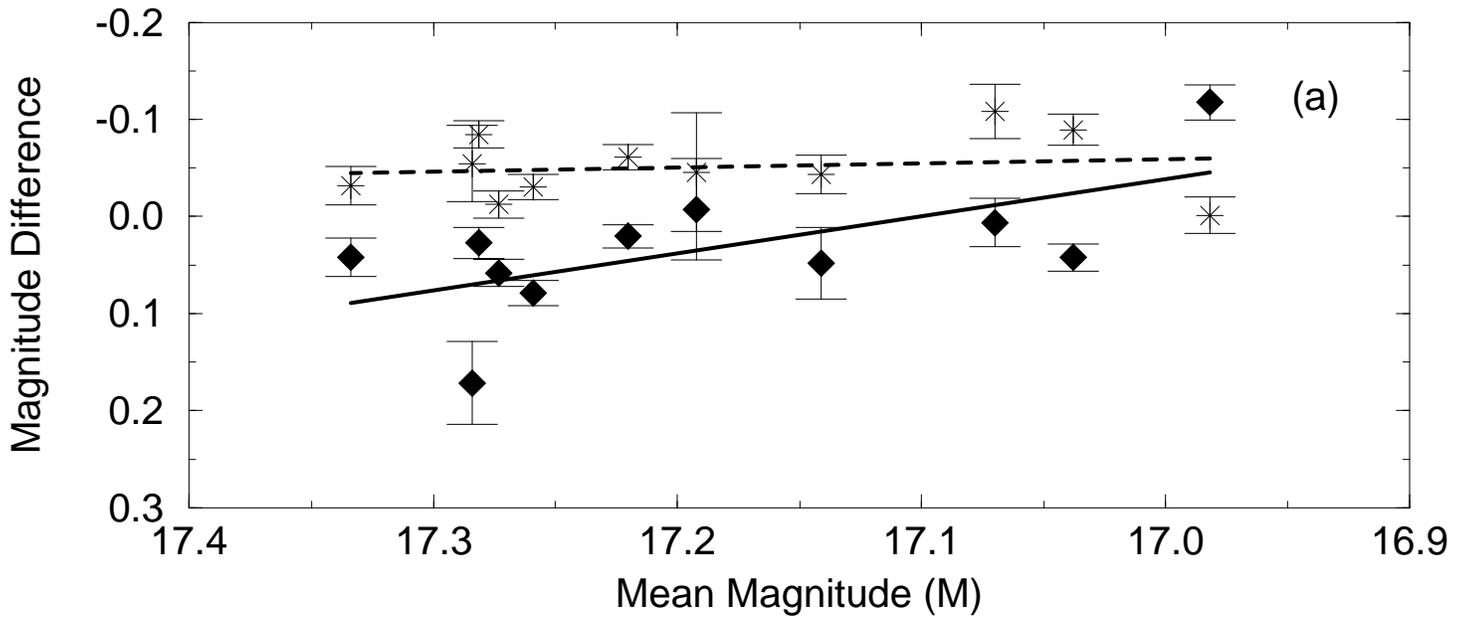
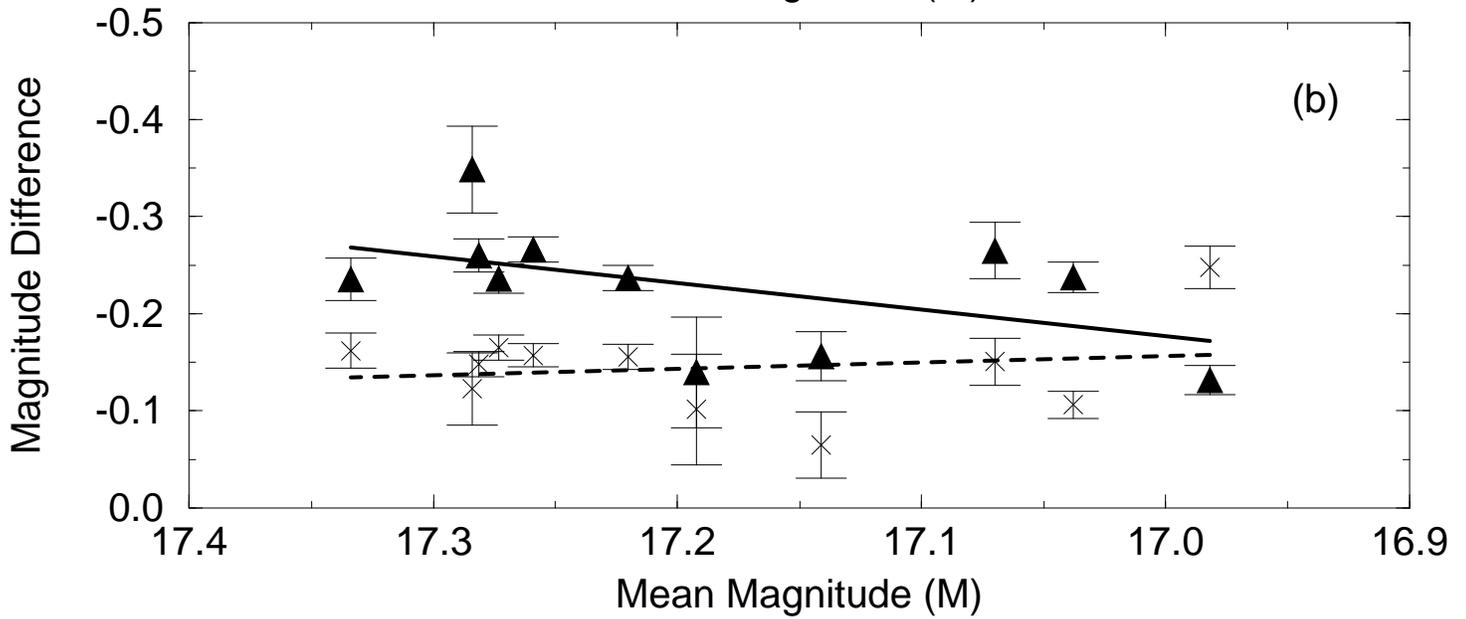
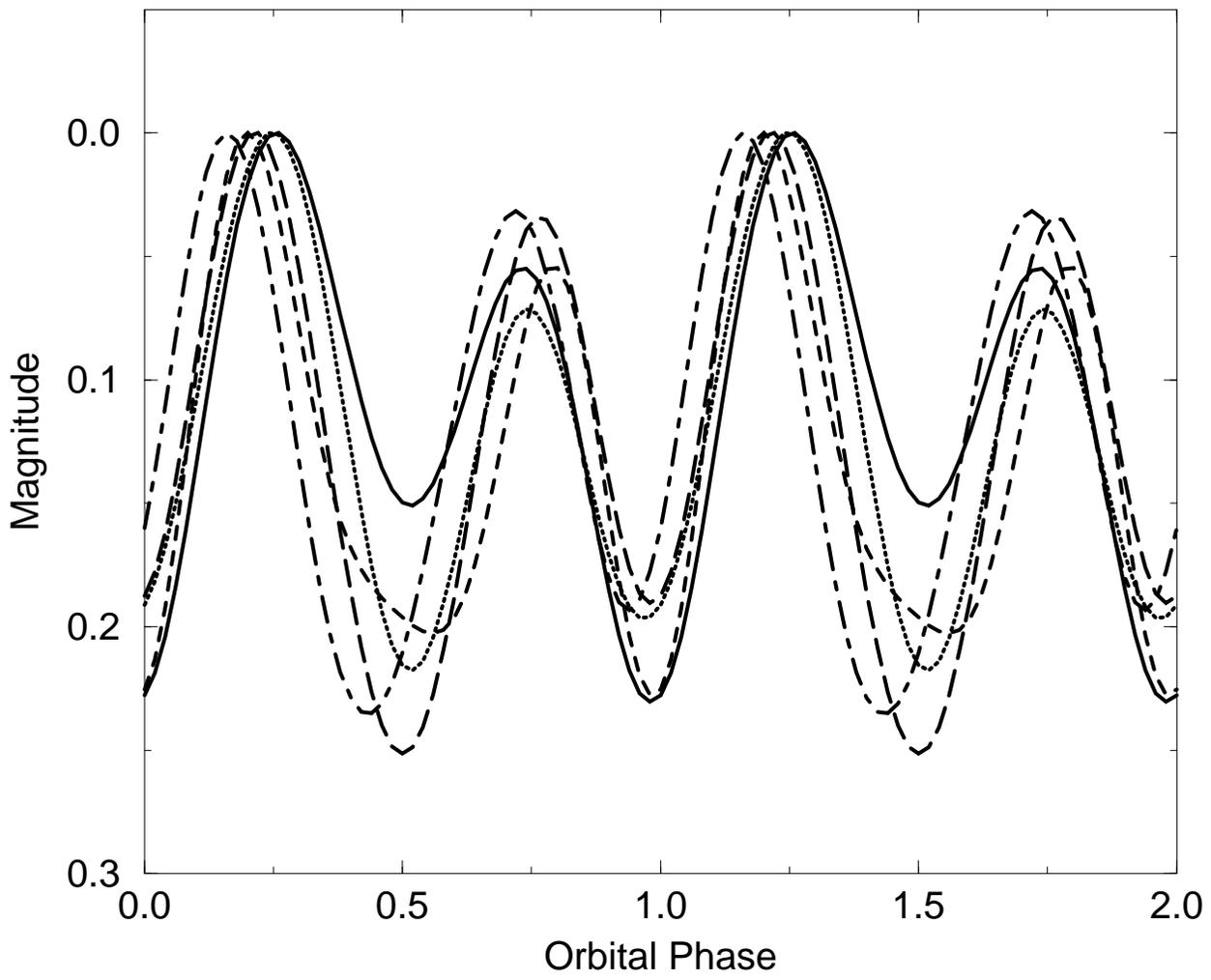
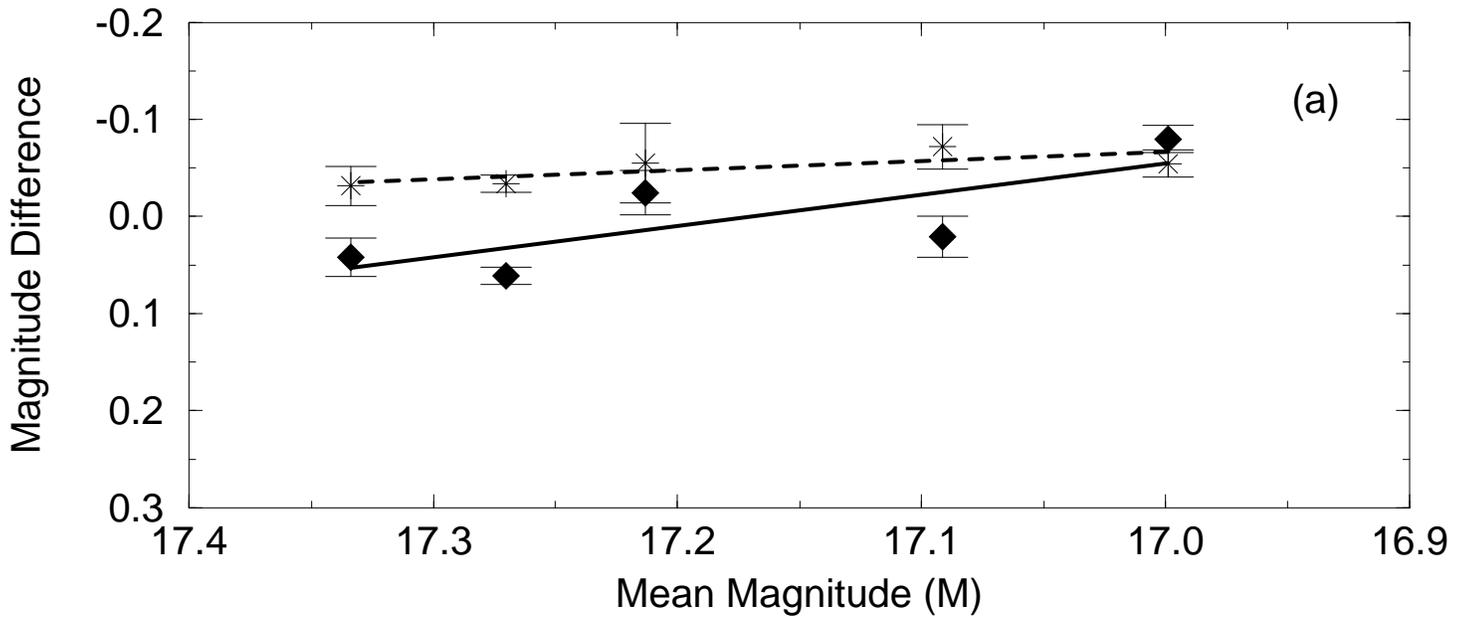
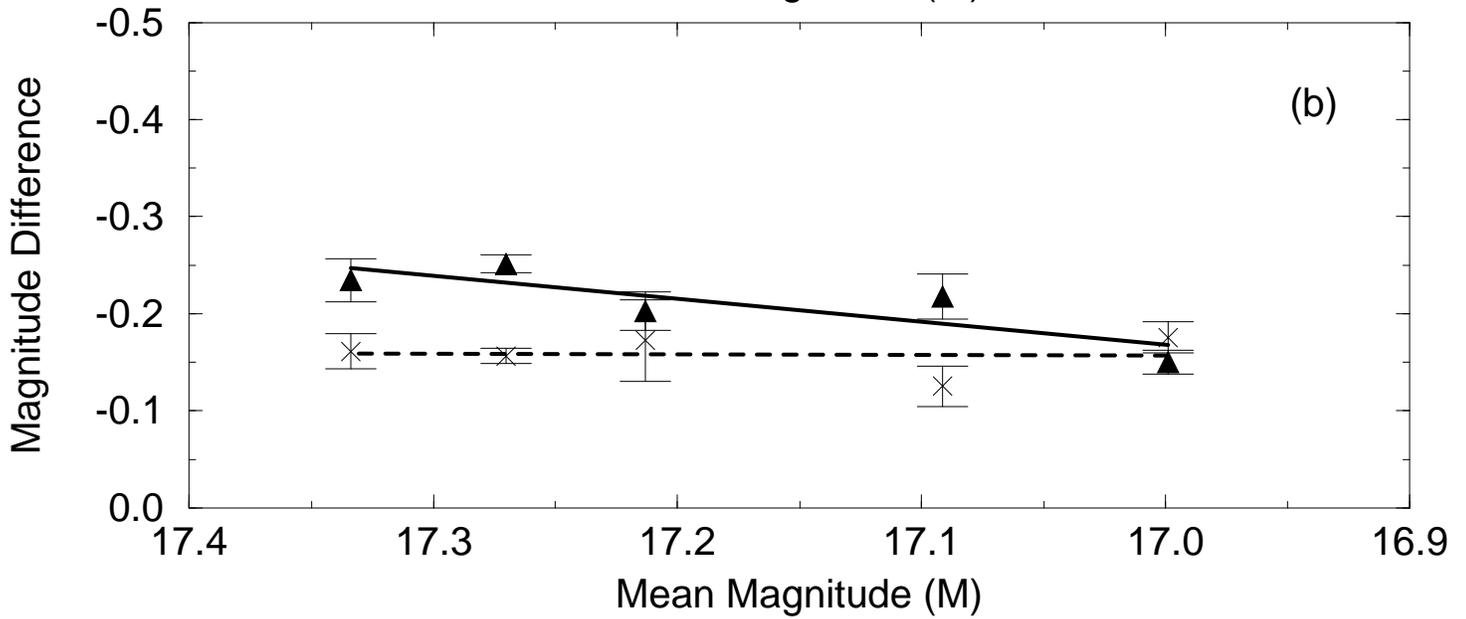
\end{document}